\documentclass[two column, superscriptaddress, pra]{revtex4-1}
\usepackage{amsfonts}
\usepackage{amsmath}
\usepackage{graphicx}
\usepackage{hyperref}
\usepackage{bbm}

\begin{document}

\title{Tuning coupling between superconducting resonators with collective qubits}
\author{Qi-Ming Chen}
\altaffiliation{Q.M.Chen is currently at Princeton University.}
\affiliation{Department of Automation, Tsinghua University, Beijing, 100084, China}

\author{Re-Bing Wu}
\email{rbwu@tsinghua.edu.cn}
\affiliation{Department of Automation, Tsinghua University, Beijing, 100084, China}
\affiliation{Beijing National Research Center for Information Science and Technology, Beijing 100084, China}

\author{Luyan Sun}
\affiliation{Center for Quantum Information, Institute for Interdisciplinary Information Sciences, Tsinghua University, Beijing 100084, China}

\author{Yu-xi Liu}
\affiliation{Beijing National Research Center for Information Science and Technology, Beijing 100084, China}
\affiliation{Institute of Microelectronics, Tsinghua University, Beijing 100084, China}

\date{\today}
\begin{abstract}
	By simultaneously coupling multiple two-level artificial atoms to two superconducting resonators, we design a quantum switch that tunes the resonator-resonator coupling strength from \emph{zero} to a large value proportional to the number of qubits. This process is implemented by engineering the qubits into different subradiant states, where the microwave photons decay from different qubits destructively interfere with each other such that the resonator-resonator coupling strength keeps stable in an open environment. Based on a three-step control scheme, we switch the coupling strength among different values within \emph{nanoseconds} without changing the transition frequency of the qubits. We also apply the quantum switch to a network of superconducting resonators, and demonstrate its potential applications in quantum simulation and quantum information storage and processing. 
\end{abstract}
\maketitle

\section{Introduction}
Superconducting circuits are of increasing importance in simulating many-body quantum physics \cite{Georgescu2014} as well as in quantum information processing \cite{Schoelkopf2008, Gu2017}. As an elementary component, coplanar stripline resonator plays a central role in microwave photon storage and transmission \cite{Blais2004, Wallraff2004}. Architectures with two \cite{Sun2006, Mariantoni2008, Reuther2010, Baust2015, Helmer2009, Johnson2010}, three \cite{Wang2011, Mariantoni2011}, and even larger arrays of superconducting resonators \cite{Raussendorf2007, DiVincenzo2009, Steffen2011, Koch2010, Underwood2012} have been reported in recent experiments. Contemporary nano-fabricating technologies enable us to couple more than a few artificial atoms and resonators in the same circuit \cite{Fink2009, Filipp2011, Loo2013, Mlynek2014, Lambert2016}.

To engineer quantum states in superconducting circuits, tunable coupling between different components is usually required in design. Quantum switches between two qubits \cite{Blais2003, Berkley2003, McDermott2005, Liu2006, Majer2007, Bialczak2011}, one qubit and one resonator \cite{Cleland2004, Sillanpaa2007, Leek2010, Eichler2012, Lu2017} (or transmission line \cite{Yin2013}), and two resonators \cite{Sun2006, Mariantoni2008, Reuther2010, Baust2015} have been proposed in the literature. Typically, the tunable resonator-resonator (R-R) interaction is realized by coupling a two-level artificial atom simultaneously to two superconducting resonators \cite{Mariantoni2008, Reuther2010, Baust2015}. When tuning the transition frequency of the qubit, the R-R coupling can be adjusted in a certain range. However, the maximum coupling strength in this design is limited by the dispersive interaction between the qubit and the two resonators. It is also unstable in an open environment because that the qubit cannot always stay at the coherent optimal point when changing the transition frequency. In these regards, we propose a new quantum switch which controls the R-R coupling in a much wider and more stable range.

The idea behind can be dated back to the \emph{1950s}, when \emph{Dicke} found that collective two-level atoms may exhibit new features that do not appear in a single atom or an ensemble of atoms with independent environments \cite{Dicke1954}. One of the fascinating phenomena is that the decoherence of the atoms can be effectively suppressed if they are prepared at the so called \emph{subradiant states} and are placed in the same environment \cite{Freedhoff1967, Stroud1972, Gross1982, Pavolini1985, Fink2009, Filipp2011}. The quantum switch proposed in this paper is realized by coupling collective artificial atoms simultaneously to two superconducting resonators, in which the R-R coupling strength can be tuned by engineering the subradiant states of the qubits while the energy and phase relaxation can be effectively suppressed. 

The rest of the paper is organized as follows. In Secs.~\ref{S:SWITCH} and \ref{S:CONTROL}, we derive the relation between the R-R coupling strength and the collective states of the qubits, and introduce a three-step method to engineer the qubits into different subradiant states. In Sec.~\ref{S:ARCHI}, we use the quantum switch to connect multiple superconducting resonators and demonstrate its possible applications in quantum simulation and quantum information. Section~\ref{S:ENVIRONMENT} studies the function of the quantum switch in an open environment and simulates the master equation of a three-resonator chain as an example. Finally, we draw conclusions and present further discussions in Sec.~\ref{S:CONCLUSION}.

\section{Theoretical description of the quantum switch}\label{S:SWITCH}
\begin{figure}
  \centering
  \includegraphics[height=6cm]{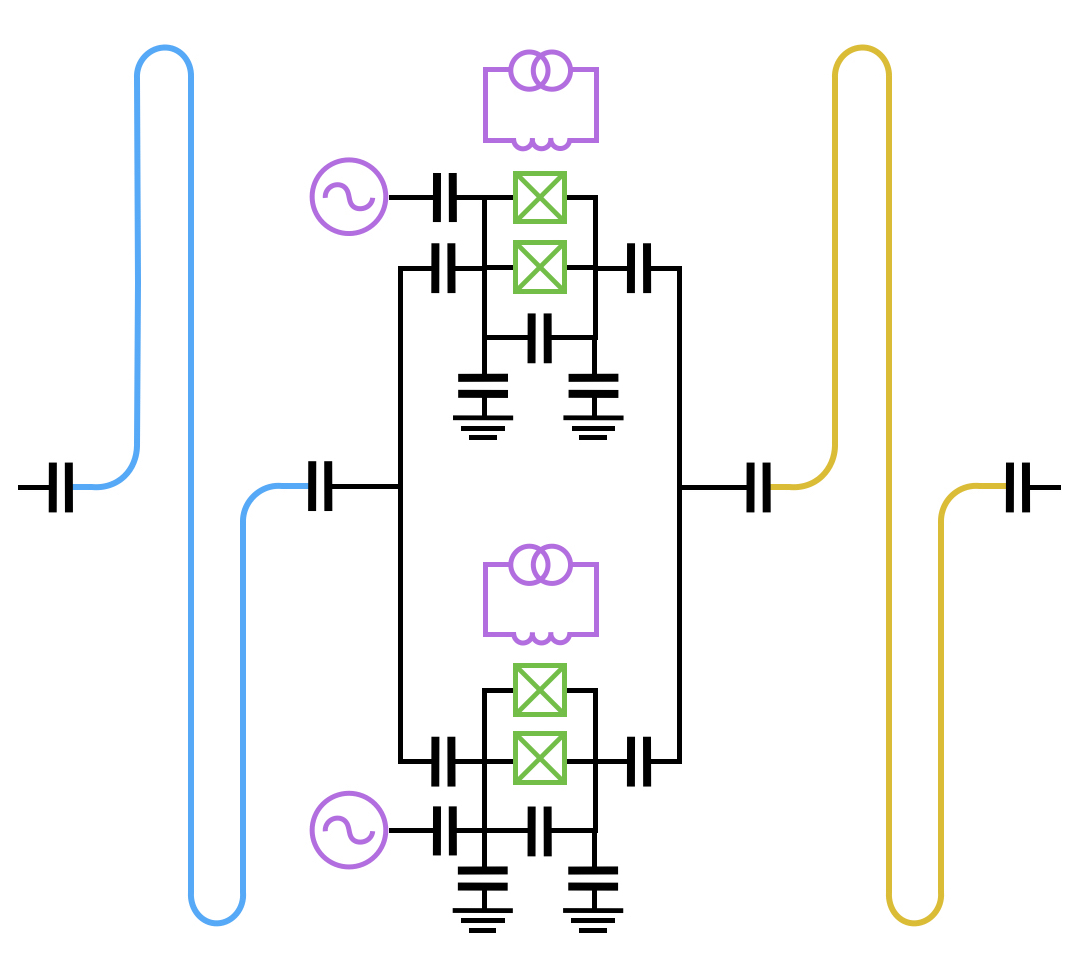}
  \caption{(Color online) Schematic diagram of the quantum switch. Two superconducting resonators A (blue) and B (yellow) are coupled through two transmon qubits (green). Transition frequency of an individual qubit can be tuned by the local superconducting coils, and direct qubit transitions can be driven by the local microwave drive lines (purple). Coupling strength between the qubits and resonators is adjusted by optimizing the relative location between the two components.}\label{F:SWITCH}
\end{figure}

As schematically shown in Fig.~\ref{F:SWITCH}, we consider a system in which two superconducting stripline resonators are coupled through $N$ superconducting artificial atoms (for example, the transmon qubits). For convenience, we assume that $N$ is an even number so that the qubits can be grouped in pairs. The case with odd number of atoms can be discussed in the same way by grouping the first $(N-1)$ atoms in pairs, and leave the last one at the ground state. As will be explained in Sec.~\ref{S:CONTROL}, we assume that the atoms can only interact with those in the same pair, such that the Hamiltonian can be written as \cite{Nigg2012}
\begin{align}
	\nonumber H^{(0)} =& \hbar \omega_a a^{\dagger}a + \hbar \omega_b b^{\dagger}b 
	+ \sum_{k=1}^{N}\hbar \omega_k c_k^{\dagger}c_k 
	+ \frac{\alpha_k}{2} c_k^{\dagger}{}^2 c_k^2 \\
	\nonumber & + \sum_{k=1}^{N}\chi_{ac_k}a^{\dagger}ac_k^{\dagger}c_k 
	+ \sum_{k=1}^{N}\chi_{bc_k}b^{\dagger}bc_k^{\dagger}c_k \\
	& + \sum_{k=1}^{N/2} K_{2k-1,2k} \left( c_{2k-1}^{\dagger}c_{2k} + c_{2k-1}c_{2k}^{\dagger} \right), 
	\label{EQ:ORI}
\end{align}
where $a$ ($a^{\dagger}$), $b$ ($b^{\dagger}$), and $c_k$ ($c_k^{\dagger}$) are the annihilation (creation) operators for the resonators A (blue), B (yellow), and the $k$th transmon qubit (green), respectively. $\alpha_k$ describes the self-Kerr interaction (nonlinerity) of the $k$th transmon mode, $\chi_{ac_k}$ ($\chi_{bc_k}$) the cross-Kerr interaction between the cavity and the transmon modes. Interaction between the two qubits in the same pair is described by $K_{2k-1,2k}$. For convenience of discussion, we assume that all the transmon qubits have the same frequency $\omega_q$, and they are equally coupled to the two resonators. This assumption can be realized by optimizing the locations of the individual qubits with respect to the two resonators \cite{Delanty2011, Loo2013}. If we reduce the transmon qubit to its lowest two levels, we obtain the standard dispersive Hamiltonian
\begin{align}\label{EQ:DISPERSIVE}
	\nonumber H^{(1)} \approx & \hbar \omega_{a} a^{\dagger}a +  \hbar \omega_{b} b^{\dagger}b 
	+ \frac{\hbar}{2}\left( \omega_q + 2\chi_{a}a^{\dagger}a + 2\chi_{b}b^{\dagger}b \right) \mathbbm{J}^z  \\
	& + \hbar g_{ab} \left( ab^{\dagger} + a^{\dagger}b \right) \mathbbm{J}^z
	+ \hbar \left( \chi_{a}+\chi_{b} \right) \mathbbm{J}^{+-}, 
\end{align}
where 
\begin{align}
    & \mathbbm{J}^z = \sum_{k=1}^{N/2}\left( \sigma_{2k-1}^z + \sigma_{2k}^z \right),\\
	& \mathbbm{J}^{+-}
	=\sum_{k=1}^{N/2}\left( \sigma_{2k-1}^{+} + \sigma_{2k}^{+} \right)
	\left( \sigma_{2k-1}^{-} + \sigma_{2k}^{-} \right). \label{EQ:COLLECT}
\end{align}
Suppose that $g_a$ ($g_b$) is the coupling strength between a single transmon qubit and the resonator A (B) \cite{Koch2007} and define $\Delta_{a(b)}=\omega_q-\omega_{a(b)}$, the coefficients in Eq.~(\ref{EQ:DISPERSIVE}) can be written as $\chi_{a} = {\left( g_{a} \right)^2 }/{\Delta_{a}}$, $\chi_{b} = {\left( g_{b} \right)^2 }/{\Delta_{b}}$, and $g_{ab}=g_ag_b(\Delta_a+\Delta_b)/(2\Delta_a\Delta_b)$. If we further define 
\begin{align}
	& \mathbbm{J}^x = \sum_{k=1}^{N/2}\left( \sigma_{2k-1}^x + \sigma_{2k}^x \right),~
	\mathbbm{J}^y = \sum_{k=1}^{N/2}\left( \sigma_{2k-1}^y + \sigma_{2k}^y \right),
\end{align}
then according to the commutation relation
\begin{equation}
	\left[ \mathbbm{J}^{i}, \mathbbm{J}^{j} \right] = i 2 \epsilon_{ijk} \mathbbm{J}^{k},~\text{($\epsilon_{ijk}$ is the \emph{Levi-Civita} symbol)}
\end{equation}
the operators $\mathbbm{J}^x$, $\mathbbm{J}^y$, and $\mathbbm{J}^{z}$ can be simply treated as angular momentum for the quantum switch, called the \emph{collective angular momentum} \cite{Sakurai1994}. In these regards, we define the eigenstates $|j,m\rangle$ of the operator $\mathbbm{J}^z$ as follows
\begin{align}
    & \mathbbm{J}^{z}| j, m \rangle = 2m | j, m \rangle, \label{EQ:SUB1} \\
	& \mathbbm{J}^{\pm}| j, m \rangle = \sqrt{(j \mp m)(j \pm m+1)} | j, m \pm 1 \rangle, \label{EQ:SUB2}
\end{align}
where $j=0,\cdots,N/2$ is related to the total angular momentum, $m=-j,\cdots,j$ is the angular momentum in the $z$ direction, $\mathbbm{J}^{\pm} = \left( \mathbbm{J}^x \pm i \mathbbm{J}^y \right)/2$ is the raising/lowing operator of all the $N$ qubis.

According to the fourth term in the dispersive Hamiltonian $H^{(1)}$, the effective coupling strength between the two resonators A and B can be varied from $-N\hbar g_{ab}$ through \emph{zero} to $N\hbar g_{ab}$ when the qubits are engineered at different collective states. In principle, the R-R coupling strength can be rather strong if the qubit number $N$ is very large and the qubits are prepared at the state $|N/2,\pm N/2\rangle$ with $m=\pm N/2$. It can also be switched off when the qubits are prepared at the state $|j,0\rangle$ with $m=0$. Thus, coupling between the two superconducting resonators can be tuned in a rather wide range by engineering the collective states of the artificial atoms. 

\section{Control of the resonator-resonator coupling}\label{S:CONTROL}
\begin{figure}
  \centering
  \includegraphics[height=6cm]{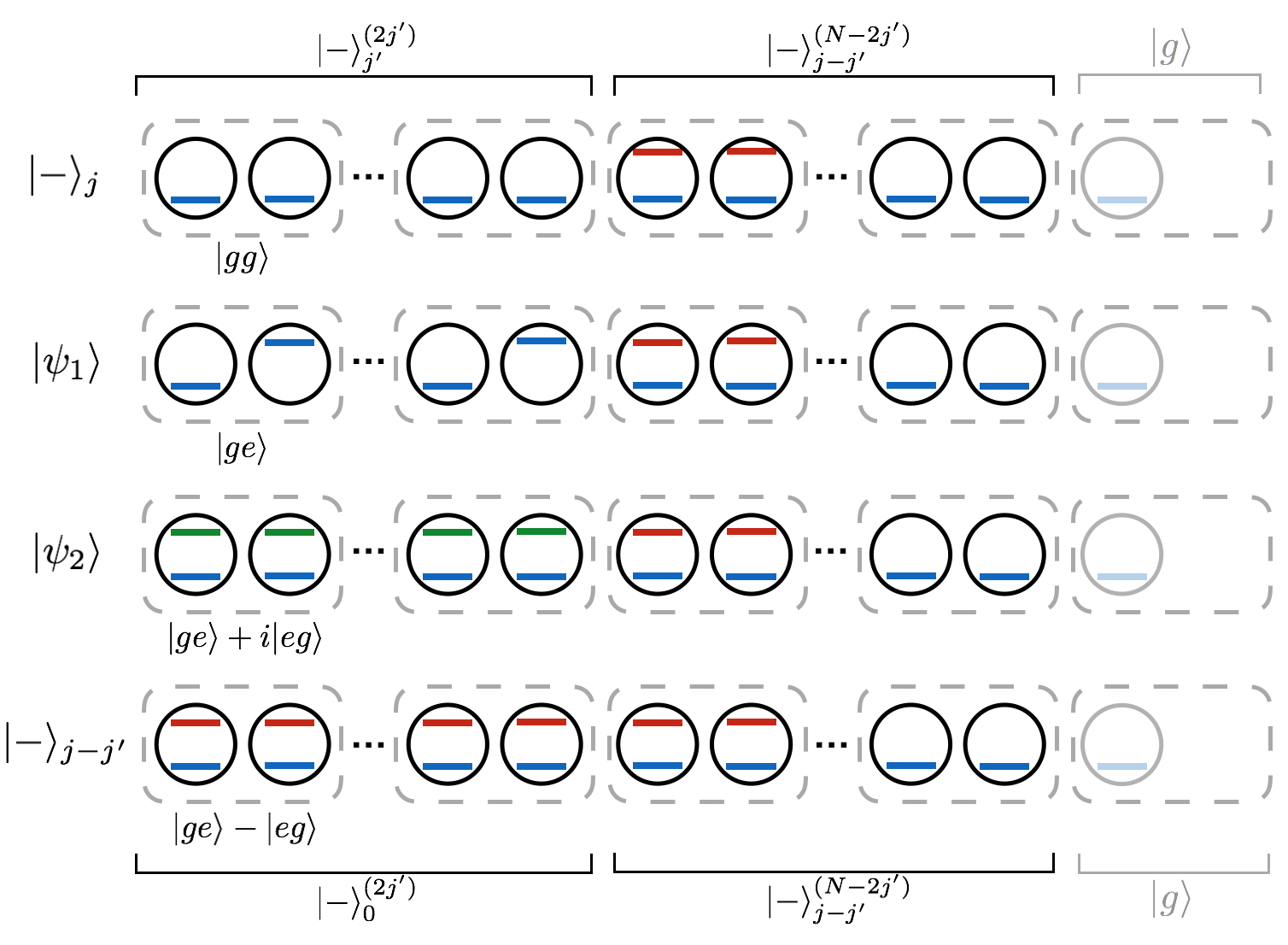}
  \caption{(Color online) Schematic diagram of the control process, where we have grouped the qubits in pairs. Started from the initial state $|-\rangle_j$, the R-R coupling strength is switched from $-2j\hbar g_{ab}$ to $-2(j-j')\hbar g_{ab}$ after the first control step, and the qubits are engineered at $|\psi_1\rangle$. The following two steps engineer the qubits within the subspace expanded by $|j-j',m\rangle$ with $m=-(j-j')$. The state $|\psi_2\rangle$ is generated after the second process, then the third step is applied to correct the relative phase in the first $j'$ qubit pairs and finally prepare the target state $|-\rangle_{j-j'}$ for the qubit collection. The light color qubit is always stayed at the ground state, which represents the case with odd number of qubits in a quantum switch. }\label{F:CON}
\end{figure}

As will be explained in Sec.~\ref{S:ENVIRONMENT}, we consider the engineering of the subradiant states of the collective qubits. The subradiant states are defined as the eigenstates of the collective angular momentum operators $\mathbbm{J}^{z}$ and $\mathbbm{J}^{-}$, i.e., 
\begin{equation}
	\mathbbm{J}^{z} |-\rangle_{j}= -2j|-\rangle_{j},~ 
	\mathbbm{J}^{-} |-\rangle_{j} = 0,
\end{equation}
where $|-\rangle_j$ is an abbreviation of the state $|j,-j\rangle$ with $m=-j$. Given $N$ qubits in the same environment, the collective angular momentum $j$ can take $N/2$ values. When $j$ varies from \emph{zero} to $N/2$, the effective R-R coupling strength varies from \emph{zero} to $-N\hbar g_{ab}$. 

Since the subradiant states are degenerate for each $j \neq N/2$, those composed of the direct product of two-qubit states is enough to describe all the possible R-R coupling strengths \cite{Scully2015}. Thus, the preparation of the subradiant states can be simplified by engineering the individual qubit pairs. To simplify the notation, we define the $2n$-qubit subradiant state with collective angular momentum $j$ as follows
\begin{equation}\label{E:NDEF}
	|-\rangle_{j}^{(2n)}= 
	\overbrace{|gg\rangle \cdots |gg\rangle}^{\text{$(n+j)$ qubits}}
	\underbrace{|\phi_{-}\rangle \cdots |\phi_{-}\rangle}_{\text{$(n-j)$ qubits}},~j=1,\cdots,n,
\end{equation}
where $n=1,\cdots,N/2$ is the number of qubit pairs, $|\phi_{-}\rangle=(|eg\rangle-|ge\rangle)/\sqrt{2}$ is the singlet state composed of two qubits. The sequence of the state $|gg\rangle$ and $|\phi_{-}\rangle$ is trivial since we can always relabel the qubit pairs and write the collective state in the form of Eq.~(\ref{E:NDEF}). Suppose that the qubits are initially prepared at the subradiant state $|-\rangle^{(N)}_{j}$ which corresponds to the effective R-R coupling strength $-2j\hbar g_{ab}$, our aim is to switch the coupling strength to $-2(j-j')\hbar g_{ab}$. 

The engineering of the subradiant states requires the multi-qubit interactions among the qubits. However, the qubit-qubit interactions are usually rather small in the dispersive regime so that the switch of the coupling strength could be very slow. Here we propose a three-step control scheme that switches the R-R coupling strength and the subradiant state separately, and thus accelerates the control of the quantum switch by a large extent. As schematically shown in Fig.~\ref{F:CON}, the first step is to flip $j'$ qubits in the first $j'$ pairs from the ground state to excited state, i.e., we engineer the first $j'$ qubit pairs from $|gg\rangle$ to $|eg\rangle$. After this procedure, the qubits are prepared at the following state
\begin{equation}
	|\psi_1\rangle = \overbrace{|eg\rangle \cdots |eg\rangle}^{\text{$2j'$ qubits}} | - \rangle_{j-j'}^{(N-2j')}.
\end{equation}
Because $|ge\rangle$ corresponds to \emph{zero} collective angular momentum in a qubit pair, the total collective angular momentum is decreased by $j'$ after the first step, i.e., the R-R coupling strength is switched from $-2j\hbar g_{ab}$ to $-2(j-j')\hbar g_{ab}$.

The following operations aim to prepare the subradiant state $|-\rangle_{j-j'}^{(N)}$ while keeping the R-R coupling strength achieved in step-one unchanged. This can be realized by taking advantage of the free evolution of the system. According to the commutation relation
\begin{equation}\label{EQ:COMMU}
	\left[ \mathbbm{J}^z, \mathbbm{J}^{+-} \right]=0,
\end{equation}
the last term in the dispersive Hamiltonian commutates with all the other terms. Thus, the free evolution of the whole system can be separated into two parts
\begin{equation}
	U(t) = U'(t) e^{-i\left( \chi_{a}+\chi_{b} \right)t \mathbbm{J}^{+-}},
\end{equation} 
where $U'(t)$ is the unitary operator generated by all the other terms in $H^{(1)}$. Since the collective angular momentum is conserved in free evolution and $|\psi_1\rangle$ is an eigenstate of $\mathbbm{J}^{z}$, the unitary operator $U'(t)$ acts only on the two oscillating modes $a$ and $b$ but not on the qubits. For the two resonators, they interact with each other according to the R-R coupling strength achieved in the first step, and the interaction is not influenced by the evolution of the qubits. For the qubits, the operator $\mathbbm{J}^{+-}$ couples only those in the same pair, i.e.,
\begin{align}
	|\psi(t)\rangle 
	=&\nonumber \mathcal{N}
	\left(|\phi_{-}\rangle + e^{-i2\left( \chi_{a}+\chi_{b} \right)t}|\phi_{+}\rangle \right) \otimes
	\cdots \\
	&\otimes \left(|\phi_{-}\rangle + e^{-i2\left( \chi_{a}+\chi_{b} \right)t}|\phi_{+}\rangle \right) | - \rangle_{j-j'}^{(N-2j')},
\end{align}
where $|\phi_{+}\rangle=(|eg\rangle+|ge\rangle)/\sqrt{2}$, $\mathcal{N}$ is a normalization factor. After a time duration $t=\pi/4\left( \chi_{a}+\chi_{b} \right)$, the qubits evolve into the following entangled state
\begin{equation}
	|\psi_2 \rangle
	= \mathcal{N}\left(|ge\rangle + i |eg\rangle \right) 
	\cdots \left(|ge\rangle +i |eg\rangle \right) | - \rangle_{j-j'}^{(N-2j')},
\end{equation}
which is shown in the third row in Fig.~\ref{F:CON}.

Finally, we apply single-qubit rotations in the first $j'$ qubit pairs to correct the relative phase between $|ge\rangle$ and $|eg\rangle$. For each qubit pair, this process is described as follows
\begin{align}
	|\psi_3\rangle_{\rm (pair)} = e^{-i \left( \sigma_z \otimes \mathbbm{1} \right) \pi/4}\left(|ge\rangle +i |eg\rangle \right) = |\phi_{-}\rangle.
\end{align}
In superconducting circuits, the single-qubit phase rotation can be implemented by applying off-resonant drives on the qubits which induce the \emph{ac-Stark} shift in the $z$-axis. Alternatively, one can use a sequence of single-qubit rotations in the $x$ and $y$ axes to construct a desired phase gate \cite{Blais2007}. 

Under typical parameters in superconducting circuits, frequencies of the resonators and the qubits are in GHz, while the coupling strength between them are MHz in the dispersive regime. Thus, the first and third steps can be implemented within \emph{nanoseconds} by using the local microwave drives (purple lines in Fig.~\ref{F:SWITCH}). The second step, by comparison, is slower since it requires the weak multi-qubit interactions to generate entanglement among the qubits. However, since the collective angular momentum is only tuned in the first step and remains invariant in the following operations, the control of the R-R coupling is implemented with only \emph{nanoseconds}.

\section{Programmable coupling among multiple resonators}\label{S:ARCHI}
\begin{figure}
  \centering
  \includegraphics[height=6cm]{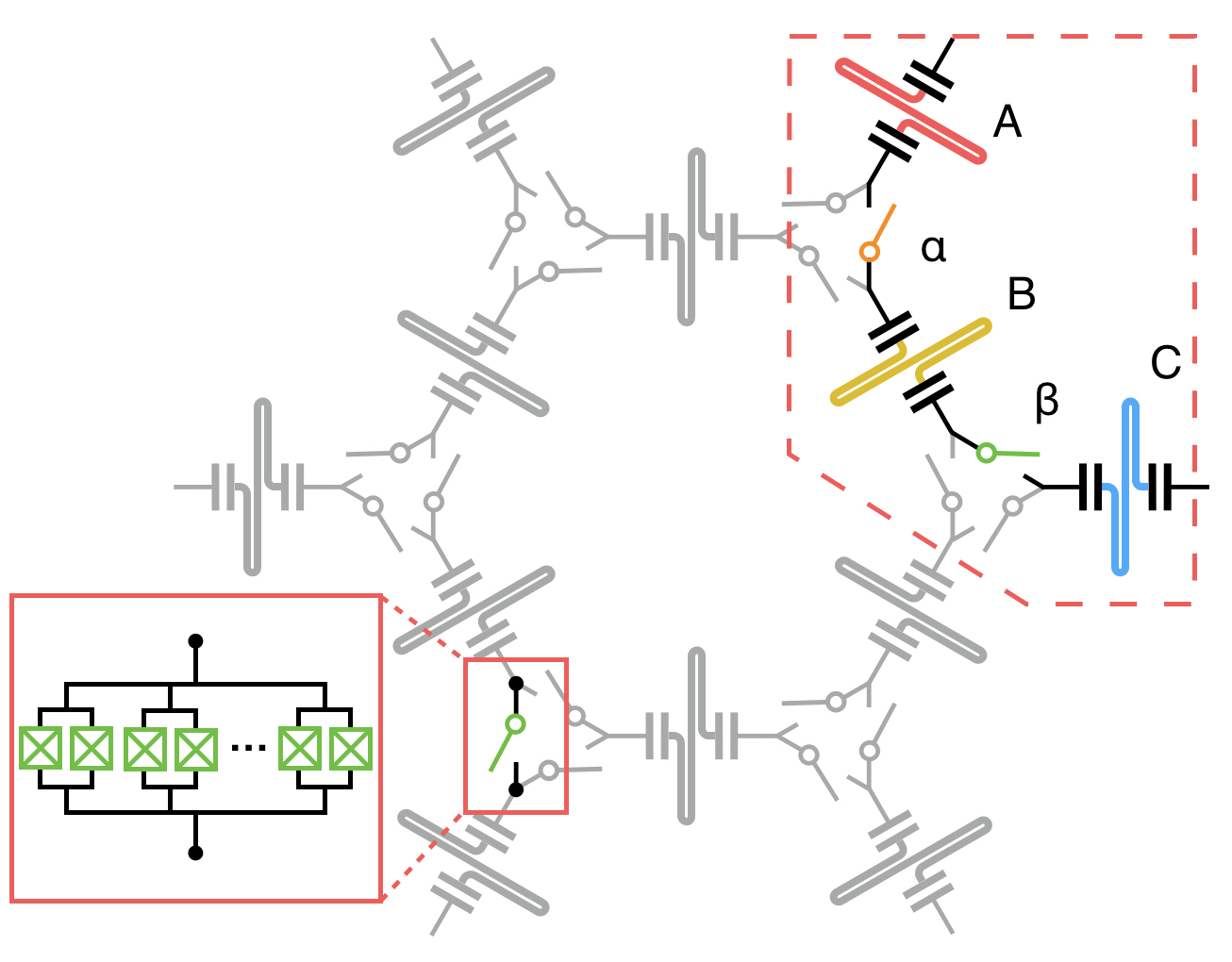}
  \caption{(Color online) Schematic diagram of the programmable coupling network for multiple superconducting resonators. The resonators are connected with each other by the quantum switches shown in Figure~\ref{F:SWITCH}, where each quantum switch is composed of $N$ transmon qubits (red solid rectangle). The interactions among different resonators can be effectively controlled by controlling the status of the quantum switches, i.e., engineering the collective qubits into different subradiant states. The red dashed rectangle shows a controlled three-resonator chain which is numerically simulated in Figure~\ref{F:SIMU}. }\label{F:ARCHITECTURE}
\end{figure}

Using the quantum switch we designed, programmable coupling among multiple superconducting resonators can be realized in an architecture illustrated in Fig.~\ref{F:ARCHITECTURE}, where the interaction between any two resonators can be controlled by the quantum switches mediated in between. Take the three-resonator chain as an example (red dashed rectangle), we study the following two types of applications of the programmable resonator network.

\subsection{Programmable quantum simulator}\label{SUB:SIMU}
Suppose that the three resonators are resonant with each other with frequency $\omega_r$, the effective Hamiltonian of the system can be derived by using the same procedure as in Eq.~(\ref{EQ:ORI})-(\ref{EQ:DISPERSIVE}), i.e., 
\begin{widetext}
\begin{align}\label{EQ:SIMU}
	\nonumber H^{(1)} \approx & \sum_{k=1}^{3}\hbar \omega_r a_k^{\dagger}a_k 
	+ \frac{\hbar}{2}\left(\omega_q^{(1)} + 2\chi_{1}^{(1)} a_1^{\dagger}a_1 + 2\chi_{2}^{(1)} a_2^{\dagger}a_2 \right) \mathbbm{J}_1^z
	+ \frac{\hbar}{2}\left(\omega_q^{(2)} + 2\chi_{2}^{(2)} a_2^{\dagger}a_2 + 2\chi_{3}^{(2)} a_3^{\dagger}a_3 \right) \mathbbm{J}_2^z \\
	&+ \hbar g_{1, 2} \left( a_1a_2^{\dagger} + a_1^{\dagger}a_2\right) \mathbbm{J}_1^z
	+ \hbar g_{2, 3} \left( a_2a_3^{\dagger} + a_2^{\dagger}a_3\right) \mathbbm{J}_2^z 
	+ \hbar\left( \chi_{1}^{(1)} + \chi_{2}^{(1)} \right) \mathbbm{J}_1^{+-} 
	+ \hbar\left( \chi_{2}^{(2)} + \chi_{3}^{(2)} \right) \mathbbm{J}_2^{+-},
\end{align}
\end{widetext}
where $a_k$ ($a_k^{\dagger}$) are the creation (annihilation) operators for the resonators A (red), B (yellow), and C (blue); $\mathbbm{J}^{z}_{1}$ ($\mathbbm{J}^{z}_{2}$) and $\mathbbm{J}^{+-}_{1}$ ($\mathbbm{J}^{+-}_{2}$) are the collective angular momentum operators for the two quantum switches $\alpha$ (orange) and $\beta$ (green). For $k=1,2$, the R-R coupling between the $k$th and $(k+1)$th resonators is described by
\begin{equation}
	g_{k,k+1} = g_k^{(k)}g_{k+1}^{(k)} (\Delta_k^{(k)}+\Delta_{k+1}^{(k)})/(2\Delta_{k}^{(k)}\Delta_{k+1}^{(k)}),
\end{equation}
 where $\Delta_{k'}^{(k)}=\omega_q^{(k')}-\omega_r$, $\chi_{k'}^{(k)} = {g_{k,k'}^2}/{\Delta_{k'}^{(k)}}$, with $g_{k}^{(1)}$, $g_{k}^{(2)}$ being the coupling strength between the the $k$th resonator and the two quantum switches.

When the two qubit collections are prepared at the subradiant states, the last three terms of in Eq.~(\ref{EQ:SIMU}) vanish and we obtain the following hopping interaction in the system
\begin{equation}
	H^{(1)}_{\rm int} = \sum_{k=1}^{2} \hbar g_{k,k+1} \left(a_{k}a_{k+1}^{\dagger} + a_{k}^{\dagger}a_{k+1}\right) \mathbbm{J}_k^z.
\end{equation}
For a more general case with $m$ resonators in a chain, interaction between any two adjacent resonators can be derived as follows
\begin{equation}
	H'^{(1)}_{\rm int} = \sum_{k=1}^{m} \hbar g_{k,k+1} \left(a_{k}a_{k+1}^{\dagger} + a_{k}^{\dagger}a_{k+1}\right) \mathbbm{J}_k^z.
\end{equation}

As demonstrated in Refs.~\cite{Koch2010, Underwood2012}, a similar architecture without quantum switches has been studied in recent researches. This resonator network is proven to be very useful in the quantum simulation of many-body physics. Different from that, the effective R-R couplings in our system can be tuned in a wide range by controlling the state of the quantum switches. The network topology of the quantum simulator is also programmable according to the status of the quantum switches. This provides more possibilities in simulating the quantum many-body physics with various interactions.

\subsection{Quantum information storage and processing}
Let us now study the non-resonant case where the two high-quality resonators A and C with the frequency $\omega_s$ are far detuned from the low-quality bus resonator B with frequency $\omega_b$. To manifest the effective interaction between the two storage resonators, we apply the following transformation on Eq.~(\ref{EQ:SIMU}) 
\begin{equation}\label{EQ:JC2}
	U_2=\exp\left[ \sum_{k=1}^{2}\frac{g_{k,k+1}}{\Delta_{sb}}\left( a_ka_{k+1}^{\dagger} - a_{k}^{\dagger}a_{k+1}\right) \mathbbm{J}_k^{z} \right],
\end{equation}
where $\Delta_{sb}=\omega_{s}-\omega_{b}$. Using the \emph{Baker-Hausdorff} lemma and omitting the higher-order terms, we obtain the following effective Hamiltonian
\begin{widetext}
\begin{align}\label{H2}
	\nonumber H^{(2)} \approx & \sum_{k=1}^3\hbar \omega_{k} a_k^{\dagger}a_k
	 +\frac{\hbar}{2}\left[ \omega_q^{(1)} + 2\left(\chi_1^{(1)} +\frac{g_{1,2}^2}{\Delta_{sb}}\mathbbm{J}_{1}^z\right)a_1^{\dagger}a_1 + 2\left(\chi_2^{(1)}-\frac{g_{1,2}^2}{\Delta_{sb}}\mathbbm{J}_{1}^z\right)a_2^{\dagger}a_2 \right] \mathbbm{J}_1^z \\
	& + \frac{\hbar}{2}\left[ \omega_q^{(2)} + 2\left(\chi_2^{(2)}-\frac{g_{2,3}^2}{\Delta_{sb}}\mathbbm{J}_{2}^z\right)a_2^{\dagger}a_2 + 2\left(\chi_3^{(2)}+\frac{g_{2,3}^2}{\Delta_{sb}}\mathbbm{J}_{2}^z\right)a_3^{\dagger}a_3 \right] \mathbbm{J}_2^z 
	+ \hbar \frac{g_{1,2}g_{2,3}}{\Delta_{sb}}\left(a_1^{\dagger }a_3+a_1 a_3^{\dagger}\right)\mathbbm{J}_{1}^{z}\mathbbm{J}_{2}^{z},
\end{align}
\end{widetext}
where or the same reason as in the derivation of  Eq.~(\ref{EQ:SIMU}), we have assumed that the qubit collections are prepared at the subradiant states. The interaction between the two storage resonators is described by the last term in Eq.~(\ref{H2}), which is proportional to the product of the collective angular momentum of the two quantum siwtches. For the $m$-resonator case with only the $1$st and the $m$th resonators being the storage resonators, the effective interaction between the two terms can be derived to be
\begin{equation}\label{EQ:INT_PRO}
	H_{\rm int}^{(2)} = \hbar g_{1,m}\left( a_1 a_{n}^{\dagger} + a_{1}^{\dagger} a_n \right) \prod_{k=1}^{m-1}\mathbbm{J}_k^{z},
\end{equation}
where 
\begin{equation}
	g_{1,m}=\prod_{k=1}^{m-1}g_{k,k+1}/{\left(\Delta_{sb}^{m-2}\right)}.
\end{equation}
For simplicity, we suppose that every qubit collection consists of $N$ artificial atoms, thus the R-R coupling strength between the two distant storage resonators can be varied from \emph{zero} to $-N^{m-1}\hbar g_{1,m}$ by controlling the quantum switches. Even if $g_{1,m}$ is usually very small in the dispersive regime, the distant R-R coupling strength can still be comparable to the nearby R-R coupling strength $g_{k,k+1}$ as long as $N\geq \Delta_{sb}/g_{k,k+1}$. Thus, this architecture not only enables quantum information storage in the individual storage resonators when any of the corresponding quantum switches are turned off, but also fast quantum information processing between two distant storage resonators when all the switches are turned on. Combined with the state preparation \cite{Law1996, Liu2004, Hofheinz2008, Hofheinz2009} and non-demolition measurement schemes \cite{Wallraff2005, Schuster2007, Johnson2010} in the literature, this provides a scalable quantum information processor with full control of quantum information read-in, storage, processing, and read-out. 

\section{The Master equation simulation in an open environment}\label{S:ENVIRONMENT}
Let us study the environmental effects on the system made of superconducting resonators and quantum switches. We assume that $\kappa_{r}$ is the decay rate of the $r$th resonator $a_r$, $\gamma_k$ and $\gamma_k'$ are the energy and phase relaxation rate of the $k$th quantum switch, the master equation for the system reads 
\begin{align}\label{EQ:MASTER}
	\nonumber \frac{d}{dt} \rho =& -\frac{i}{\hbar} [ H^{(0)}, \rho ] 
	+ \sum_r \kappa_{r}\mathcal{D} \left[ a_r \right] \\
	& + \sum_k \frac{\gamma_k}{2} \mathcal{D} \left[ J_k^{-} \right]
	+ \sum_k \frac{\gamma_k'}{2} \mathcal{D} \left[ J_k^{z} \right],
\end{align}
where 
\begin{align}
 J_k^{z} = \sigma_{2k-1}^{z} + \sigma_{2k}^{z},~
 J_k^{\pm} = \sigma_{2k-1}^{\pm} + \sigma_{2k}^{\pm},	
\end{align}
are two-qubit operators of the $k$th qubit pair, $\mathcal{D}\left[ X \right] = 2 X \rho X^{\dagger} - X^{\dagger}X \rho - \rho X^{\dagger}X$ is the \emph{Lindblad} superoperator. We have assumed in Eq.~(\ref{EQ:MASTER}) that the two qubits in the same pair share the same environment, which has been reported in recent experiments  \cite{Fink2009, Filipp2011}. Since the subradiant state we have discussed in Sec.~\ref{S:CONTROL} is a direct product of the two-qubit singlet state $|\phi_{-}\rangle$ and the ground state $|gg\rangle$, the collective states of the artificial atoms are also the eigenstates of the \emph{Lindblad} superoperators. Thus, the status of the quantum switches are not influenced by the energy and phase relaxation in an open environment, and the R-R coupling controlled by the quantum switch is rather stable in an open environment. 

\begin{figure}
  \centering
  \includegraphics[height=8.5cm]{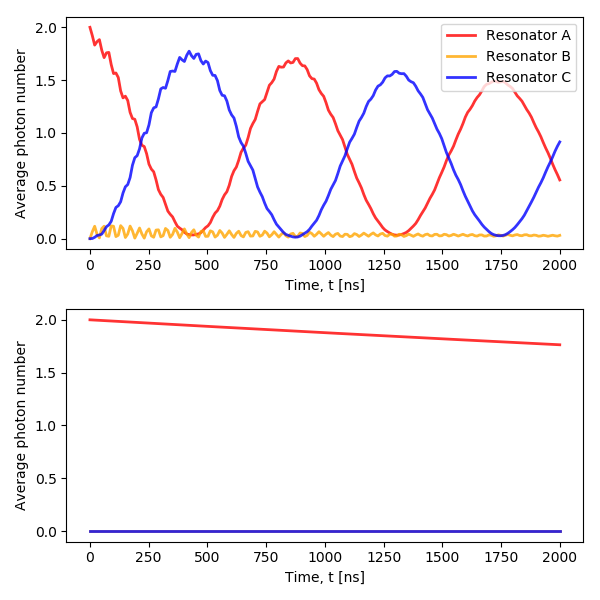}
  \caption{(Color online) Master equation simulation of the three-resonator chain presented in the dashed rectangle in Figure~\ref{F:ARCHITECTURE}. The two storage resonators A (red) and C (blue) exchange microwave photons with each other when the both quantum switches are turned on, while state of the bus resonator B (yellow) is not influenced over time (upper). This interaction can be effectively turned off by switching off one of the quantum switches (lower).  Parameters of the system are chosen in the typical parameter regime of transmon qubit and the transmission line resonators. In detail, $\omega_s/2\pi = 5.18~{\rm GHz}$, $\omega_b/2\pi = 5.20~{\rm GHz}$, $\omega_{q}/2\pi = 5.00~{\rm GHz}$, $g_s^{(1),(2)}/2\pi = 18.0~{\rm MHz}$, $g_b^{(1),(2)}/2\pi = 20.0~{\rm MHz}$, $\kappa_s/2\pi = 5~{\rm MHz}$, $\kappa_b/2\pi = 200~{\rm MHz}$, $\gamma_{1,2}/2\pi = 20~{\rm MHz}$, $\gamma'_{1,2}/2\pi = 20~{\rm MHz}$.}\label{F:SIMU}
\end{figure}

Figure~\ref{F:SIMU} simulates the three-resonator chain shown in the dashed rectangle in Fig.~\ref{F:ARCHITECTURE}, where the two quantum switches are both turned on (upper), or the quantum switch $\alpha$ is turned off (lower) \cite{Johansson2013}. For simplicity, we define $g_s$, $\Delta_s$ ($g_b$, $\Delta_b$) as the coupling strength and frequency difference between a storage (bus) resonator and a qubit in the quantum switch. The initial state of the three resonators are prepared at $|2\rangle$, $|0\rangle$, and $|0\rangle$, and the parameters are chosen to meet the requirement of dispersive coupling $\Delta_{s} (\Delta_{b}) \sim 10g_{s} (10g_{b})$, $\Delta_{sb}\sim 10g_{1,2} (10g_{2,3})$. As expected, the exchange of microwave photons between the two storage resonators can be effectively controlled by the quantum switch, i.e., by engineering the subradiant states of the two qubit collections. When the two switches are turned on, the coupling strength between the two storage resonators achieves $\sim 0.4g_{1,2}$ with $N=2$. This coupling strength can be further enhanced by increasing the number of qubits in the quantum switch. For example, the distant coupling reaches $\sim 0.9g_{1,2}$ when $N=3$, which is comparable to the coupling strength between two adjacent resonators.

As we can see, the \emph{Rabi} flopping between the two storage resonators decays with time, which is mainly caused by the unwanted coupling to the low-quality bus resonator. This phenomenon shows a trade-off between increasing the R-R coupling ratio $g_{1,3}$ and increasing the frequency difference between the bus and the storage resonators $\Delta_{sb}$. However, because that the R-R coupling is quadratic to the qubit number $N$ but only linear to the frequency difference $\Delta_{sb}$, the relaxation can be effectively suppressed by increasing the frequency detuning while the coupling strength maintained or even enhanced by increasing the qubit number. 

When the quantum switche $\alpha$ is turned off, the storage resonators A and C cannot interact with each other so that the photons are stored in the individual resonators. However, photon number in A still decays slowly with time due to the environmental effects. Compared with the decay rate of a bare storage resonator, the photon leakage we have observed is mostly caused by the intrinsic decay of the resonator $\kappa_s$. The quantum switch does not influence the life time of the microwave photons stored in the storage resonator, so that the the quantum information is well protected in the individual resonators. 

\section{Conclusions and discussions}\label{S:CONCLUSION}
To conclude, we propose a quantum switch that can tune the coupling strength between two superconducting resonators in a wide range. This switch consists of a collection of two-level artificial atoms that are coupled simultaneously to two superconducting resonators. By preparing the qubits at different states with different collective angular momentum,  the resonator-resonator coupling strength varies from \emph{zero} to $N\hbar g$ which is proportional to the number of qubits $N$ (suppose $g$ is the coupling strength between the qubits and the resonator). Consider the dissipation of the qubits in an open environment, we group the qubits in pairs and propose a three-step control scheme to engineer the qubits into different subradiant states. These subradiant states are free of decoherence by assuming that the two qubits in each pair share the same environment, and the coupling strength can be switched within \emph{nanoseconds}.

In addition, we use the quantum switch to connect multiple resonators with a programmable network topology, in which the interaction between any two resonators can be adjusted \emph{in situ}. According to the situation whether the resonators are on resonant or largely detuned from each other, this resonator network can be used to simulate various quantum many-body physics in superconducting circuits, or store and process quantum information in a scalable manner. 

Although transmon qubits are addressed in this paper, other types of superconducting qubits such as charge, flux, and phase qubits can also be used with little modification. Natural qubits such as neutral atoms, ions, and spins should also be available. This type of hybrid quantum system would benefit from the long coherence time of the natural qubits as well as the uniform coupling between the resonators and the qubits, and thus results in a more reliable quantum switch than that designed in this paper. However, this proposal may be limited by the weak coupling strength between the natural qubits and the superconducting resonators \cite{Xiang2013}, and it will be studied elsewhere. 

\section*{Acknowledgement}
R.B.W acknowledges the support of the National Natural Science Foundation of China (Grant No. 61773232, 61374091, and 61134008) and National Key Research and Development Program of China (Grant No. 2017YFA0304300). LS acknowledges the support of the National Natural Science Foundation of China Grant No.11474177 and 1000 Youth Fellowship program in China. Y.X.L. acknowledges the support of the National Basic Research Program of China Grant No. 2014CB921401 and the National Natural Science Foundation of China Grant No. 91321208.
 
\appendix
 
\bibliography{REF}  
\end{document}